# Deep Assessment of Code Review Generation Approaches: Beyond Lexical Similarity

YANJIE JIANG, Peking University, China

HUI LIU, Beijing Institute of Technology, China

TIANYI CHEN, Beijing Institute of Technology, China

FU FAN, Beijing Institute of Technology, China

CHUNHAO DONG, Beijing Institute of Technology, China

KUI LIU, Huawei Software Engineering Application Technology Lab, China

LU ZHANG, Peking University, China

Code review is a standard practice for ensuring the quality of software projects, and recent research has focused extensively on automated code review. While significant advancements have been made in generating code reviews, the automated assessment of these reviews remains less explored, with existing approaches and metrics often proving inaccurate. Current metrics, such as BLEU, primarily rely on lexical similarity between generated and reference reviews. However, such metrics tend to underestimate reviews that articulate the expected issues in ways different from the references. In this paper, we explore how semantic similarity between generated and reference reviews can enhance the automated assessment of code reviews. We first present a benchmark called GradedReviews, which is constructed by collecting real-world code reviews from open-source projects, generating reviews using state-of-the-art approaches, and manually assessing their quality. We then evaluate existing metrics for code review assessment using this benchmark, revealing their limitations. To address these limitations, we propose two novel semantic-based approaches for assessing code reviews. The first approach involves converting both the generated review and its reference into digital vectors using a deep learning model and then measuring their semantic similarity through Cosine similarity. The second approach generates a prompt based on the generated review and its reference, submits this prompt to ChatGPT, and requests ChatGPT to rate the generated review according to explicitly defined criteria. Our evaluation on the GradedReviews benchmark indicates that the proposed semantic-based approaches significantly outperform existing state-of-the-art metrics in assessing generated code review, improving the correlation coefficient between the resulting scores and human scores from 0.22 to 0.47.

**ACM Reference Format:**

---

Authors' Contact Information: Yanjie Jiang, yanjiejiang@pku.edu.cn, Peking University, China; Hui Liu, liuhui08@bit.edu.cn, Beijing Institute of Technology, China; Tianyi Chen, chentianyiyx@gmail.com, Beijing Institute of Technology, China; Fu Fan, fufan@bit.edu.cn, Beijing Institute of Technology, China; Chunhao Dong, dongchunhao22@bit.edu.cn, Beijing Institute of Technology, China; Kui Liu, brucekuiliu@gmail.com, Huawei Software Engineering Application Technology Lab, China; Lu Zhang, zhanglucs@pku.edu.cn, Peking University, China.

---

## 1 Introduction

Code review is a widely adopted practice crucial for ensuring and enhancing the quality of source code [5,6,38]. Typically, a pull request—comprising a set of changes to a software project—can only be merged into version control systems if it has been reviewed and approved by designated reviewers. While code reviews are often time-consuming, they provide significant benefits by enabling human experts to suggest improvements and preventing the integration of unqualified changes. To mitigate the high cost of manual code reviews, there has been significant recent research into automated code review generation. For instance, Tufano et al. [39] employ the pre-trained model Text-To-Text Transfer Transformer to generate code review comments automatically without human intervention.

Despite significant advancements in the automated generation of code reviews, the automated assessment of code review generation approaches remains less explored [41]. A rigorous and comprehensive evaluation of these approaches is crucial for several reasons. Firstly, any new approach or software system, including those for code review generation, must be thoroughly assessed before deployment in industry settings. Secondly, to advance the state of the art, researchers need to pinpoint the inaccuracies of existing methods, which requires extensive evaluation and comparison. Finally, new approaches need to be rigorously and quantitatively evaluated against current standards to gain the confidence of reviewers and potential users.

A straightforward method for assessing code review generation approaches involves applying these methods to a set of pull requests and manually evaluating the quality of the generated reviews. The main advantage of manual assessment is its high accuracy and reliability, provided the evaluators are qualified. However, a significant drawback is the resource-intensive nature of manual evaluation. To address this, automated performance metrics are used to assess the quality of automatically generated code reviews. One common approach is to measure the lexical similarity between generated reviews and reference reviews (i.e., reviews created by human experts). The most frequently used metric for this purpose is BLEU [22], which measures the similarity between an automatically generated review and its reference, a method borrowed from machine translation. Another metric, known as Exact Match [40], assesses how often the generated reviews are lexically identical to their references. These metrics offer objective and easy-to-compute assessments, making the evaluation and comparison of code review generation approaches both convenient and objective.

While existing performance metrics such as BLEU and Exact Match have contributed significantly to the automated assessment of code review generation approaches, they often lack accuracy and fail to fully capture the similarity between generated and reference reviews as suggested by the evaluation results in Section 6. Ideally, if a generated review conveys the same or substantially similar information (comments and/or suggestions) as the reference review, it should be considered similar or a high-quality review. However, both BLEU and Exact Match rely solely on lexical similarity, completely overlooking the semantic relationship between reviews. As a result, these metrics tend to significantly underestimate reviews that present the expected issues differently from the references. Although these reviews convey the same information, their use of different words or phrases results in low lexical similarity, despite their high semantic equivalence.

To enhance the automated assessment of code review generation approaches, we explore how semantic similarity between generated reviews and reference reviews can be leveraged for this task. This investigation requires a benchmark in which generated code reviews have been manually evaluated. However, such high-quality benchmarks are currently lacking. To address this gap, we introduce a benchmark called GradedReviews, which is built by collecting real-world code reviews from open-source applications, generating code reviews using state-of-the-art approaches, and manually grading the quality of these generated reviews. The final benchmark consists of 5,164 automatically generated code reviews, each paired with human-assigned scores.

Table 1. Motivating Examples

| | Reference Review | Generated Review | BLEU | Embedding based Similarity | LLM Score | Human Score |
|---|---|---|---|---|---|---|
| 1 | We don't need super here | Unnecessary call to super | 17.53 | 0.512 | 4 | 4 |
| 2 | why waste time whitelisting it? | why do you want to whitelist it at the end? | 12.88 | 0.636 | 4 | 4 |
| 3 | swallow? | stringbuilder? | 70.71 | 0.219 | 1 | 1 |

Using this benchmark, we evaluate the performance of existing state-of-the-art metrics for code review assessment, revealing the quantitative limitations of these metrics. We then propose two semantics-based approaches for assessing code reviews. The first approach, embeddingSim, transforms the generated review and its reference into digital vectors using a deep embedding model, and measures their semantic similarity through Cosine similarity between the vectors. The second approach, LLM-based Scoring, generates a prompt based on the generated review and its reference, feeds this prompt into ChatGPT, and requests ChatGPT to rate the generated review according to explicitly defined criteria. Our evaluation on the benchmark indicate that both proposed approaches significantly outperform BLEU, the current state-of-the-art metric for code review assessment.

The paper makes the following contributions:

- A **benchmark** for code review evaluation where all reviews have been manually scored.
- An **empirical study** that reveals the limitations of the state-of-the-art performance metrics for code review generation approaches.
- Two simple yet effective **approaches** to assessing code review generation approaches.

## 2 Motivating Examples

In this section, we present real-world examples, shown in Table 1, to demonstrate why lexical similarity-based performance metrics may fail to accurately measure the similarity between code reviews. The first example involves the code review: "We don't need super here", addressing the source code snippet "super();". The reviewer suggests that calling super is unnecessary in this context. For the same pull request, Tufano's tool [39], a state-of-the-art code review generation approach, generates the review: "Unnecessary call to super". This generated review correctly identifies the issue—an unnecessary call to super—making it highly relevant and useful, earning a score of 4 (excellent) from human evaluators. However, widely-used performance metrics such as BLEU rate the generated review as poor because it shares only a single token ("super") with the reference review, resulting in a BLEU score of just 17.53. This example highlightshow lexical similarity-based metrics can significantly underestimate the quality of generated code reviews. Another example involves the generated review: "Why do you want to whitelist it at the end?", which is lexically different from its reference review: "Why waste time whitelisting it?". As a result, the BLEU score is as low as 12.88, labeling it as poor based on lexical similarity. However, both reviews convey the same essential point: the whitelist is unnecessary. The generated review earns a high score of 4 (excellent) from human evaluators despite the low BLEU score.

Lexical similarity-based metrics like BLEU not only underestimate certain reviews but can also overestimate others. A typical example is the code review generated by commentFinder[15]: "stringbuilder?", which is fundamentally different from the reference review "swallow?". The term "swallow" typically refers to the practice of catching an exception or error and then silently ignoring it, without taking any meaningful action or providing feedback. The reference review's intent is to suggest that the code should take meaningful actions when handling the caught exception. In contrast, the generated review "stringbuilder?" is entirely unrelated to this intent, and thus received a low score of 1 (poor) from human evaluators. Despite this, the BLEU score for the generated review is up to 70.71, as half of the tokens in both reviews overlap. This example demonstrates how lexical similarity-based metrics can significantly overestimate the quality of a poor review.

From Table 1, we observe that the scores generated by large language models (LLMs) are highly consistent with human evaluations. By feeding the generated and reference reviews into ChatGPT and requesting it to grade the generated reviews based on their reference reviews, the resulting scores, as shown in Table 1, are identical to those assigned by human evaluators. This suggests that LLMs have the potential to assess automatically generated code reviews with much greater accuracy than the widely used BLEU metric.

Additionally, the table shows that embedding-based similarity is more accurate than BLEU for evaluating generated code reviews. Our approach converts the reviews into digital vectors using deep embedding models, and then computes their semantic similarity using Cosine similarity between the vectors. For example, the two high-quality reviews in Table 1 exhibit strong embedding-based similarity with their reference reviews, with scores of 0.512 and 0.636, respectively. In contrast, the poor review—the last entry in the table—has a much lower embedding-based similarity score of 0.219 with its reference review. In conclusion, embedding-based similarity demonstrates a strong alignment with human evaluations, a consistency that BLEU lacks.

Based on the preceding examples, we conclude that lexical similarity-based metrics, such as BLEU, are inadequate for accurately assessing the quality of generated code reviews in terms of their alignment with corresponding reference reviews. This observation serves as the primary motivation for the research presented in this paper.

## 3   Related Work

### 3.1   Code Review Generation

Several approaches have been proposed for automatically generating code reviews. Early methods primarily relied on retrieval-based techniques to extract relevant historical reviews. For instance, Gupta et al. [13] introduced DeepMem, which recommends code reviews for code snippets by first searching a repository for candidate reviews and then using an LSTM model to select the most relevant one. Siow et al. [33] enhanced it by incorporating attention mechanisms into LSTM models, enabling a deeper semantic understanding of the code and its corresponding reviews. Yang et al. [15] proposed CommentFinder, which calculates the cosine similarity between the method under review and previously revised methods. Based on this similarity, it selects the ten closest matches and then applies Gestalt Pattern Matching (GPM) [1] to identify the most similar method. The associated code review is returned as the recommended review.

To the best of our knowledge, Tufano et al. [39,40] were the first to generate code reviews using deep learning models. They employed a pre-trained Text-To-Text Transfer Transformer (T5) to generate code reviews for given source code. Li et al. [22] proposed CodeReviewer, a pre-trained model designed specifically for code review generation. It incorporates four pre-training tasks tailored for the code review process, including the generation of review comments. AUGER [20] differs from other deep learning-based approaches by utilizing an automated algorithm to explicitly link code review to the specific lines of code they address.

Large language models have also been applied to code review generation. Lu et al. [24] introduced an innovative framework, LLaMA-Reviewer, to recommend code reviews using the LLaMA large language model [37]. CodeMentor proposed by Nashaat et al. [27] is the first in this line to exploit reinforcement learning with human feedback from domain experts. They first prepares organizational data to build a domain-specific model, fine-tunes the model with instruction-based data, and then refine the model further through reinforcement learning. Tufano et al. [38] conducted an empirical study investigating the capabilities of three state-of-the-art techniques: commentFinder [15], CodeReviewer [22], and a pre-trained model-based approach [39]. They also explored ChatGPT for code reviews generation. Zhou et al. [44] empirically evaluated existing generation-based automatic code review techniques along with general-purpose pre-trained code models. Their results demonstrate that CodeT5 frequently achieves the best performance in review generation.

Table 2. Datasets and Performance Metrics for Evaluating Code Review Generation Approaches

| Review Generation Approaches | Datasets | Performance Metrics | References |
|---|---|---|---|
| DeepMem (2018) | Gupta's | MRR, Exact Match | [13] |
| Core (2020) | Siow's | MRR, Exact Match | [33] |
| Tufano's Tool (2022) | Tufano's | BLEU, Exact Match | [39] |
| CodeReviewer (2022) | Zhiyu Li's | BLEU | [22] |
| Auger (2022) | Lingwei Li's | ROUGE | [20] |
| CommentFinder (2022) | Tufano's | BLEU, Exact Match | [15] |
| LLaMA-Reviewer (2023) | Tufano's, Zhiyu Li's | BLEU | [24] |
| Zhou's Tool (2023) | Tufano's | Exact Match | [44] |
| CodeMentor (2024) | Zhiyu Li's | BLEU | [27] |

## 3.2 Benchmark for Code Review Generation

To evaluate various code review generation approaches, different benchmarks and metrics have been developed, as summarized in Table 2. Gupta et al. [13] trained and tested DeepMem on C# code reviews collected from 208 representative code repositories in Microsoft. This benchmark contains 22,435 completed C# pull requests and 56,052 code-and-review pairs. Siow et al. [33] randomly selected 19 projects from GitHub's top 200 repositories, ranked by the number of stars. Based on these projects, they collected 57,260 <code change, review> pairs. Tufano et al. [39] mined Java open-source projects and extracted triplets < $m_s$, $c_{nl}$, $m_r$ >, where $m_s$ is the method submitted for review, $c_{nl}$ is the reviewer's comment for $m_s$, and $m_r$ is the revised version of $m_s$ implementing the reviewer's recommendations. Li et al. [22] mined open-source projects on GitHub, covering the nine most popular programming languages: C, C++, C#, Go, Java, JavaScript, PHP, Python, and Ruby. In total, they collected 10,168 reviews. Similarly, Li et al. [20] selected 11 Java repositories from GitHub, compiling 79,344 code reviews.

Our benchmark differs from the datasets mentioned above. While these datasets were designed for the evaluation or training of code review generation approaches, our benchmark is specifically constructed to assess various performance metrics for code review generation approaches. Another key distinction is that the reviews in our benchmark are accompanied by human-assigned scores, a feature absent from the other datasets.

## 3.3 Performance Metrics for Code Review Generation

Various performance metrics have been proposed to evaluate code review generation approaches, including Exact Match [40], BLEU [28], MRR [25], and ROUGE [23].

Exact Match [40] measures how often the generated reviews are identical to the reference reviews. It is typically used for approaches that produce a single review for a given input. When multiple reviews can be generated for a single input, Mean Reciprocal Rank (MRR) [12,25,42], a standard metric in information retrieval, can be employed. MRR calculates the rank of the correct review, where a review is considered correct only if it matches the reference exactly.

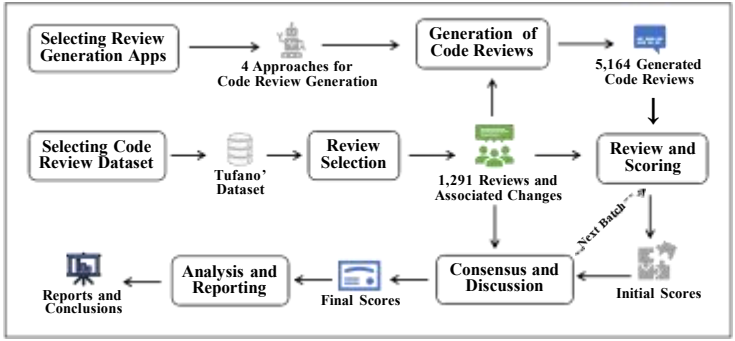

Fig. 1. Methodology for Benchmark Construction

Bilingual Evaluation Understudy (BLEU) [28], a widely-used metric for assessing machine-generated translations, is often applied to generated code reviews. It measures lexical similarity between the generated review and the reference review based on n-grams [36]. Similarly, ROUGE [23], another popular metric from machine translation, has been adapted for code review evaluation. ROUGE captures the overlap of words and the longest matching sequence between the generated review and its reference.

Our metrics differ from these traditional approaches as we are the first to introduce semantics-based metrics in this domain, whereas existing metrics focus on lexical similarity between reviews.

## 4 Benchmark Construction

We observe that existing datasets of code reviews typically lack manual scores, particularly for generated reviews. This omission makes it difficult to assess the reliability of performance metrics like BLEU in evaluating the quality of generated code reviews. To address this gap, we construct a large dataset of code reviews, each accompanied by manual scores, allowing other researchers to evaluate the effectiveness of performance metrics for code review generation. Fig.1 outlines the methodology used for constructing the benchmark, with further details provided in the following subsections.

### 4.1 Selection of Dataset and Reference Code Reviews

We select Tufano's dataset [39] for benchmark construction for several key reasons. First, it is one of the most widely used datasets for code review generation, and the majority of papers published in this field after its release have used it for evaluation[15,24,44]. Utilizing such a widely adopted dataset facilitates the code review generation process, a critical step in our benchmark construction as outlined in Section 4.2. Second, the dataset is large and of high quality, consisting of 167,799 reviews from open-source projects, each item is a triplet, including the method submitted for the review, a single reviewer's comment, and a revised version of the method. Additionally, the dataset is explicitly divided into a training set (denoted as Trset) and a testing set (denoted as Teset). Since most state-of-the-art code review generation tools have been trained on Trset, we leverage only the testing set, Teset, for our benchmark construction. This approach helps mitigate the risk of data contamination.

We observe that the selected dataset, Teset, contains 16,780 reference code reviews. Given the high cost and labor-intensive nature of manually scoring each generated review, we adopt a sampling approach to reduce the workload while maintaining the diversity of the resulting benchmark. Specifically, we sample one item for every 13, yielding a set of 1,291 reference reviews. Importantly, instead of applying true random sampling, we use a pseudo-random method by systematically selecting positions (e.g., 1st, 14th, 28th, and so on). This pseudo-random sampling ensures that the selection process is fully reproducible.

The resulting dataset, referred to as seeds, comprises 1,291 reference reviews, each item with the associated method code. To facilitate mapping between seeds and Teset, we retain the original IDs from Teset for each item in seeds.

## 4.2  Selection of Tools and Generation of Code Reviews

After surveying automated code generation methods, we selected four approaches for our benchmark construction: the approach proposed by Tufano et al.[39] (referred to as Tufano's), commentFinder [15], Auger [20], and LLaMA-Reviewer [24]. These approaches were chosen for the following reasons. First, they represent the current state-of-the-art in the field and were all released recently. Second, they are publicly available, which facilitates third-party evaluation. Finally, all four approaches (except for commentFinder that does not request training) have been trained using the same dataset, Trset, and evaluated using the same dataset Teset, which is a superset of seeds. Consequently, we did not need to retrain these approaches; instead, we simply downloaded the code reviews they generated for the code changes in seeds. This process resulted in a total of 5,164 reviews (4 approaches × 1,291 reviews), with each reference review in seeds being associated with four generated code reviews.

The resulting dataset, referred to as RawReoiews, consists of 1,291 code changes and 5,164 code reviews corresponding to these changes. Each code change is associated with one reference code review, created manually by expert reviewers, as well as four code reviews generated by four state-of-the-art review generation tools.

## 4.3  Manual Scoring

Manual scoring involves evaluating the quality of automatically generated code reviews by comparing them to their corresponding reference reviews. Specifically, if a generated code review exactly matches its reference review, it receives the highest score, indicating perfect quality. Importantly, we do not assess the associated code changes, assuming that the reference reviews are accurate. This assumption is consistent with the approach used in existing code review-related performance metrics, such as BLEU and Exact Match, which are computed by comparing the generated reviews directly to their reference reviews, rather than evaluating the code changes themselves.

To facilitate manual scoring, we use a five-point grading scale instead of asking participants to provide a continuous score between 0 and 100. The advantage of the five-point scale is that it allows us to define explicit criteria for each grade, which is not feasible with continuous scores. These clearly defined criteria enhance both the quality of the scoring and the consistency among multiple graders. To establish the scoring criteria, we conducted a pilot study with 20 randomly sampled machine-generated code reviews from RawReoiews. Three authors manually graded these reviews, focusing on criteria that justified the assigned scores and the differences among reviews. They reached a consensus on the following scoring criteria:

- 5 points (Perfect): The generated review is identical to the reference review.
- 4 points (Excellent): The generated review is essentially equivalent to the reference review, though the wording differs.
- 3 points (Good): The generated review accurately and explicitly addresses some comments or suggestions found in the reference review.

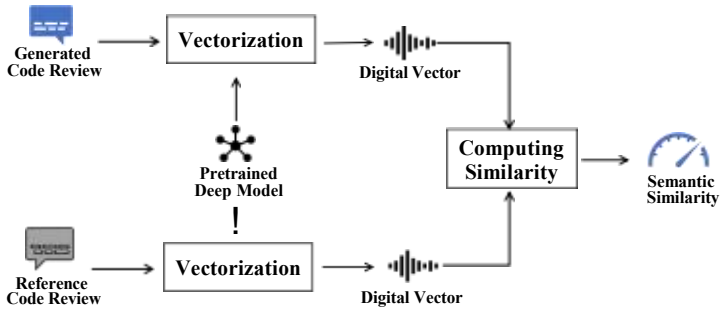

Fig. 2. Embedding-based Scoring

- 2 points (Fair): The generated review is only loosely related to the reference review.
- 1 point (Poor): The generated review is unrelated to the reference review.

Using these criteria, three authors manually scored all automatically generated reviews in RawReoiews. To minimize inconsistencies and reduce costs, we employed a batch labeling approach, dividing the reviewing tasks into three equally sized batches. Within each batch, the participants independently scored all reviews and then discussed any discrepancies to reach a consensus on each review. This process allowed them to proceed to the next batch with a deeper and more consistent understanding of the scoring criteria, which helped reduce inconsistencies and lower the cost of manual scoring. The Fleiss' Kappa coefficient [18] among the three raters improved from 0.66 (for the first batch) to 0.71 (for the second batch) and 0.76 (for the final batch).

## 4.4 Resulting Benchmark

The resulting benchmark, referred to as GradedReviews, consists of 5,164 generated code reviews, each accompanied by a manually assigned score, a reference code review, and the associated code changes to be reviewed. Notably, the distribution of scores is neither even nor normal, with a predominance of low scores. Approximately ninety percent of the scores are 1 (the lowest score), while only 0.76% (39 out of 5,164) are 5 (the highest score). This skewed distribution indicates that most automatically generated code reviews are not comparable to manually created reviews.

## 5 Semantics-based Assessment of Code Reviews

The examples provided in Section 2 demonstrate that relying solely on lexical similarity to compare code reviews is often inaccurate. Instead, a semantic comparison is crucial. To address this, we propose two semantic-based approaches for comparing code reviews. The first approach, termed embeddingSim, aligns with state-of-the-art text comparison methods by converting code reviews into digital vectors using deep embedding models and calculating semantic similarity through the Cosine similarity between these vectors. The second approach, LLM-based Scoring, represents a departure from traditional methods. Rather than using digital embeddings, it utilizes ChatGPT, a leading large language model, to directly assess the consistency between the generated review and its reference review. Detailed descriptions of embeddingSim and LLM-based Scoring are provided in Section 5.1 and Section 5.2, respectively.

## 5.1 Embedding-Based Scoring

Fig. 2 provides an overview of the embedding-based approach, embeddingSim. For a code review $r$ and its reference review $r'$, embeddingSim first employs a pre-trained deep model to convert them into equally sized digital vectors. Notably, to process texts in natural languages and feed them into deep neural networks, researchers have developed various deep models for text vectorization[31]. Compared to traditional one-hot encoding, a major advantage of these deep models is their ability to capture the semantics of the text. This means that semantically similar texts are typically represented by similar vectors[8,17], making the resulting vectors sensitive to semantic meaning. The most commonly used deep models for this purpose are described as follows:

- **CodeBERT**: It is a pre-trained model specifically designed for programming tasks [10]. CodeBERT was developed by Microsoft Research in November 2020. It is essentially an extension of BERT, optimized for both natural language and programming language understanding. CodeBERT has been pre-trained on a large corpus including both natural languages (e.g., comments and documentation) and programming languages (i.e., source code). Including both natural languages and programming languages in the pre-training enables CodeBERT to understand the relationship between code and its corresponding descriptions. This ability is crucial for software engineering tasks like code generation, and automated code review.

- **text-embedding-3-large**: It is a state-of-the-art text embedding model developed by OpenAI in 2024 [3]. It is designed to generate embeddings that represent textual data in a high-dimensional vector space. Text-embedding-3-large contains a large number of parameters, making it capable of capturing more complex relationships in the data compared to smaller models. This increased capacity generally leads to better performance on a wide range of tasks.

- **all-mpnet-base-v2**: It is a transformer-based sentence embedding model from the sentence transformers library, designed to convert sentences or short text sequences into dense vector representations [19]. Built on the MPNet [34] (Masked and Permuted Pre-training for Language Understanding) architecture, it offers improved performance over earlier BERT-based models. This model is optimized for a wide range of tasks, including semantic search and sentence similarity, and is known for its balance between efficiency and accuracy.

- **mxbai-embed-large-v1**: It is designed to generate high-quality dense vector representations of text, optimized for tasks such as sentence similarity and clustering [21]. This model can capture nuanced language features and context at a more granular level, leading to improved performance in tasks requiring a deep understanding of text.

- **UAE-Large-V1**: It is a powerful language model tailored for producing high-quality text embeddings [29]. As a large model, it excels in capturing intricate language features and contextual relationships within text, making it highly effective for tasks such as semantic similarity, information retrieval and text classification. The model's architecture allows for a deeper understanding of linguistic nuances, offering enhanced performance in scenarios that demand comprehensive text comprehension.

The generated code review r and its corresponding reference review $r^{'}$ are converted into equally sized vectors o and $o^{'}$, respectively. The semantic similarity between the r and $r^{'}$ is computed as follows:

$$\text{Cosine\_Similarity}(r, r^{'}) = \frac{\mathbf{v} \cdot \mathbf{v'}}{\|\mathbf{v}\|\|\mathbf{v'}\|} \tag{1}$$

## 5.2 LLM-based Scoring

Deep embedding models have been shown to effectively capture semantic relationships among textual data [11,16], and Cosine similarity has proven to be effective in measuring the similarity between vectors [26,35]. However, deep embedding models embed different texts (e.g., the generated code review and its corresponding reference review) independently. As a result, these models may sometimes overlook direct correlations between the independently embedded texts.

To address this, we propose a more direct and straightforward approach, termed LLM-based Scoring, for assessing a generated code review against its reference. Specifically, we input both reviews into a large language model and request it to rate the quality of the generated review by comparing it to the reference review. A key insight of the approach is that advanced language models, such as ChatGPT, excel in understanding natural language and are well-suited to comprehending and comparing two short texts, such as code reviews. Another key insight is that the model can view both the generated review and the reference review simultaneously, which facilitates direct comparison between the reviews.

A key aspect of LLM-based Scoring is its scoring system. Traditional embedding-based similarity computation often results in continuous scores ranging from zero to one. This continuous scoring represents a regression task, where the model must predict a value on a continuous scale. This requires fine-tuned calibration and an understanding of subtle differences between inputs, which can be challenging for a model like ChatGPT that was primarily designed for generating and interpreting text. In contrast, a 1-5 scoring system can be framed as a classification problem, where ChatGPT predicts a discrete category (i.e., 1, 2, 3, 4, or 5). Classification tasks are typically easier for models like ChatGPT because they involve selecting from a limited set of predefined options. Additionally, a 1-5 scoring system aligns well with the manual scoring approach described in Section 4.3, thereby simplifying the comparison between LLM-based scoring and manual scoring.

Another key consideration in LLM-based Scoring is how to guide the large language model in scoring. One approach is to fine-tune the model using a dataset of manually labeled examples, allowing it to learn scoring rules automatically from this fine-tuning data. This learning-based method avoids the need for explicit guidance or prior knowledge, making it practical in many situations. However, it requires a large, high-quality dataset with accurate scores and does not guarantee that the model will fully learn the scoring criteria. Alternatively, scoring rules can be explicitly specified in the prompts provided to the model. When such rules are available, this approach can be effective without the need for fine-tuning or a dataset for training. In our case, we have clearly defined the scoring rules in Section 4.3 for human experts. As a result, we opt for the latter approach, specifying the scoring rules directly in the prompts.

Based on the design outlined above, we propose the following prompt template:

> **System**: You are a smart code reviewer. You will be asked to grade a generated code review. You can mimic answering them in the background 10 times and provide me with the most frequently appearing answer. Furthermore, please strictly adhere to the output format specified in the question. There is no need to explain your answer.
> **Scenario Matching**: I am going to give you a generated code review as well as its reference review. You should grade the generated review by comparing it to the reference review, and output a grade based on the following criteria:
> 1. If the generated review is identical to the reference review, Grade=5;
> 2. If the generated review is essential equivalent to the reference review although their expressions are not identical, Grade=4;
> 3. If the generated review explicitly and correctly specifies some comments/suggestions presented in the reference review, Grade=3;
> 4. If the generated review is only loosely related to the reference review, Grade=2;
> 5. If the generated review is completely unrelated to the reference review in semantics, Grade=1.
> Please NOTE that you should only output a grade without any explanation.
> **Generated Code Review**: "[generated-review]"
> **Reference Code Review**: "[reference-review]"

where "generated-review" and "reference-review" should be replaced with the concrete reviews.

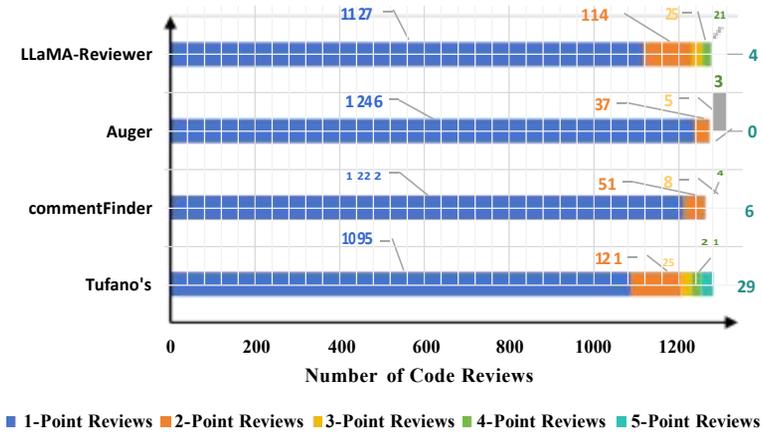

Fig. 3. Quality of Code Reviews Generated by Different Approaches

Notably, large language models like ChatGPT often suffer from randomness [4,43]. To mitigate this randomness, we implement the following measures:

- First, we directly compare the generated code review with the reference review to check for exact matches. If they are identical, we assign a grade of 5 without invoking ChatGPT.

- Second, if ChatGPT is used to grade the review, we perform the grading process three times and select the most frequently given grade. If each trial results in a different grade, we choose the median value. For example, if ChatGPT provides grades of 5, 4, and 3, we assign a final grade of 4.

- Third, if ChatGPT assigns a grade of 5 (perfect), we perform a lexical comparison to verify the equivalence between the generated review and the reference review. If they are not lexically identical, we adjust the grade to 4 (excellent) to reflect that the generated review does not meet the standard of perfection.

- Finally, if ChatGPT produces a grade outside the 1 to 5 range, we request a new grade. If it fails to provide a valid grade after three attempts, we terminate the process and do not assign a grade.

## 6 Empirical Study

### 6.1 Research Questions

- **RQ1**: How accurate are current state-of-the-art approaches to automated code review generation?
- **RQ2**: How effective is BLEU in the automated assessment of generated code reviews?
- **RQ3**: Is embeddingSim more accurate than BLEU in the automated assessment of generated code reviews?
- **RQ4**: Is LLM-based Scoring more accurate than BLEU and embeddingSim in the automated assessment of generated code reviews?

### 6.2 RQ1: State of the Art in Code Review Generation

The resulting benchmark constructed in Section 4 comprises 1,291 reference code reviews, along with 4 × 1,291 automatically generated reviews from four state-of-the-art code review generation approaches. Each generated review has been manually graded by comparing it to its reference review, as detailed in Section 4.3. The resulting grades, shown in Fig. 3, reflect the performance of the code review generation approaches.

From this figure, we make the following observations:

- **Quality of Generated Reviews**: Most generated code reviews are of low quality. Specifically, 90.82% (4,690 out of 5,164) of the generated reviews are graded as "unrelated to the reference review" (i.e., grade = 1). The second largest category is grade 2, which constitutes 6.25% (323 out of 5,164) of the reviews. According to the scoring criteria, these reviews are "only loosely related to the reference review". Together, these two lowest categories account for 97.07% (90.82% + 6.25%) of the generated reviews, indicating that most of the reviews are significantly different from their references.

- **Potential for High-Quality Reviews**: Despite the challenges in generating high-quality code reviews, all evaluated approaches produced some high-grade reviews. In total, 88 out of 5,164 reviews received a grade of 5 ("identical to the reference review") or 4 ("essentially equivalent to the reference review"), suggesting that automated code generation has notable potential.

- **Effectiveness of Approaches**: Among the four evaluated approaches, Tufano's approach proves to be the most effective. It generated 29 "perfect" reviews (grade = 5), compared to 6, 0, and 4 perfect reviews from the other three approaches, respectively. Additionally, Tufano's approach produced the fewest low-grade reviews, generating 1,095 point-1 reviews, whereas the other approaches produced 1,222, 1,246, and 1,127 point-1 reviews, respectively.

Based on the preceding analysis, we conclude that code review generation remains a challenging task and requires significant improvement.

## 6.3 RQ2: BLEU versus Human Scores

BLEU is one of the most well-recognized performance metrics used to assess generated code reviews [39]. To evaluate its effectiveness, we compare BLEU scores with human scores. First, we compute the correlation coefficients between BLEU and human scores. The Spearman Rank Correlation Coefficient of 0.22 with a p-value significantly less than 0.05 indicates a weak positive correlation between BLEU and human scores. This positive correlation suggests that higher BLEU scores are somewhat associated with higher human scores. However, the correlation is weak, indicating that a high BLEU score does not necessarily guarantee a high human score. In other words, BLEU is only a weak indicator of review quality. Accordingly, we also compute two additional commonly used metrics: ROUGE and METEOR. These metrics are chosen to complement our analysis, providing a more comprehensive assessment of the results. The Spearman Rank Correlation Coefficients for GOUGE and METEOR are 0.25 and 0.21, respectively, with a p-value significantly below 0.05. Therefore, the conclusions drawn from these two metrics are consistent with those of BLEU.

Second, we present the distribution[1] of BLEU scores for reviews with different scores in Fig. 4. Each bean in the graph represents the BLEU score distribution for reviews with a specific score, such as 1 point. Notably, we do not include a bean for 5-point reviews in the graph because all such reviews have a BLEU score of 100. This uniform distribution results in error reported by the underlying drawing algorithm: "sample is too sparse to find TD". The distribution of ROUGE and MENTER is similar to that of BLEU, the details are presented in Fig. 5 and Fig. 6. From these figures, we make the following observations:

---

[1]BLEU, ROUGE, MENTER scores are non-negative. The negative values appearing in Bean Plots are typically caused by the smoothing process of kernel density estimation.

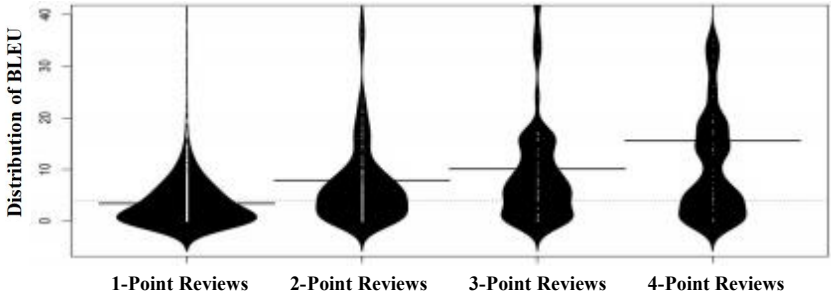

Fig. 4. Distribution of BLEU for Each Human Score

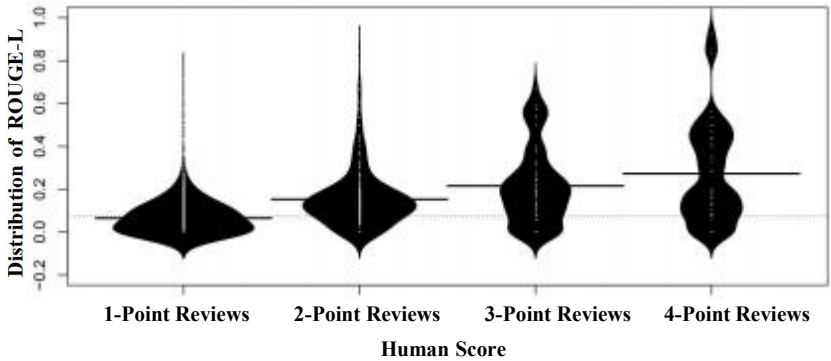

Fig. 5. Distribution of ROUGE for Each Human Score

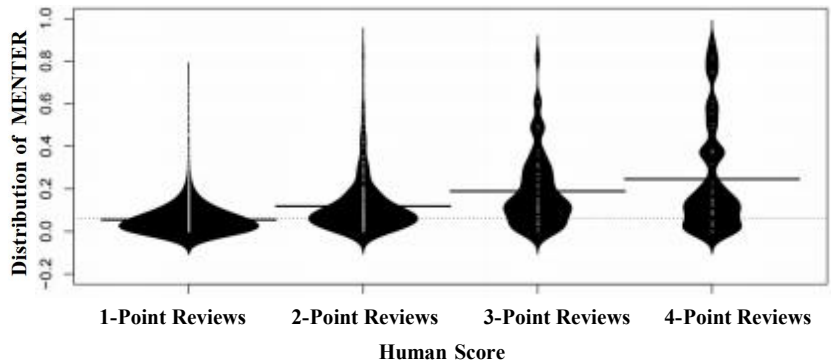

Fig. 6. Distribution of METEOR for Each Human Score

- **Median Metrics Scores:** The median BLEU score increases with the increase of human scores. Specifically, the medians for reviews graded 1, 2, 3, and 4 are 1.94, 5.12, 6.95, and 8.62, respectively. It confirms the weak positive correlation between BLEU and human scores. The median ROUGE score also increases with the increase of human scores. Specifically, the medians for reviews graded 1, 2, 3, and 4 are 0.06, 0.13, 0.19, and 0.20, respectively. The median MENTER score follows a similar trend, with medians for reviews graded 1, 2, 3, and 4 being 0.04, 0.08, 0.14, and 0.15, respectively.

Table 3. Comparison among Text Embedding Models

| Embedding Models | Spearman Rank Correlation Coefficient | Accuracy |
|---|---|---|
| text-embedding-3-large | 0.38 | 67.0% |
| all-mpnet-base-v2 | 0.34 | 57.1% |
| mxbai-embed-large-v1 | 0.32 | 56.0% |
| UAE-Large-V1 | 0.32 | 53.0% |
| CodeBert | 0.12 | 34.4% |

Table 4. Performance of Embedding-based Decision Tree

| Classes | (Macro) Precision | (Macro) Recall | (Macro) $F_1$ |
|---|---|---|---|
| 1-Point Reviews | 96% | 71% | 82% |
| 2-Point Reviews | 7% | 20% | 11% |
| 3-Point Reviews | 7% | 45% | 12% |
| 4-Point Reviews | 8% | 43% | 14% |
| 5-Point Reviews | 94% | 100% | 97% |
| All Reviews | 43% | 56% | 43% |

- **Overlap in Metric Score Distributions:** There is a significant overlap in BLEU score distributions for reviews of different scores. For instance, the BLEU scores for 2-point reviews range from 8.23e-81 to 70.33, which significantly overlaps with the range of BLEU scores for 1-point reviews (from 2.65e-104 to 70.71) and the range for 3-point reviews (from 2.95e-77 to 48.11). There is a significant overlap in ROUGE score distributions for reviews of different scores. For instance, the ROUGE scores for 2-point reviews range from 0 to 0.84, which significantly overlaps with the range of ROUGE scores for 1-point reviews (from 0 to 0.71) and the range for 3-point reviews (from 0 to 0.67). Similarly, the MENTER score distributions show considerable overlap across different review scores. The MENTER scores for 2-point reviews range from 0 to 0.85, overlapping with the range of MENTER scores for 1-point reviews (from 0 to 0.69) and the range for 3-point reviews (from 0 to 0.82). This considerable overlap makes it challenging, if not impossible, to accurately infer human scores based solely on BLEU scores.

### 6.4 RQ3: Embedding-based Scoring

Given the diversity of text embedding models, we evaluate five state-of-the-art models and compare their performance in Table 3. The comparison is based on two criteria: the strength of the correlation between the embedding-based similarity and human scores, and the accuracy of the similarity-based classifier that translates embedding-based similarity into human scores. Notably, we omit the p-values from the table, as all are significantly below the 0.05 threshold. From Table 3, text-embedding-3-large (referred to as textEmbedding3 for brevity in the remainder of this paper) outperforms the other models, exhibiting the highest correlation coefficient and accuracy. For conciseness, we will use textEmbedding3 as the representative embedding model to explore the advantages and limitations of embedding-based similarity in the subsequent sections.

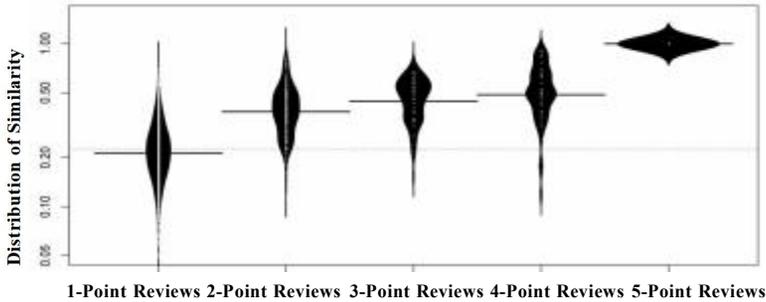

Fig. 7. Distribution of Embedding-based Similarity for Each Human Grade

When comparing embedding-based similarity to BLEU, we observe that the embedding-based similarity shows a significantly stronger correlation with human scores. The Spearman Rank Correlation Coefficient increases from 0.22 (BLEU) to 0.38 (embedding), reflecting a relative improvement of 72.7% = (0.38 - 0.22) / 0.22. Notably, a coefficient of 0.38 indicates a moderate positive monotonic relationship between embedding-based similarity and human scores, whereas a coefficient of 0.22 suggests only a weak relationship between BLEU and human evaluations.

The embedding-based similarity outperforms BLEU in its ability to be accurately mapped to human scores. The accuracy of mapping embedding-based similarity to human scores is 67%, higher than BLEU's 62.5%, yielding a relative improvement of 7.2% = (67% - 62.5%) / 62.5%. A comparison of Table 4 with Table ?? reveals that embedding-based mapping substantially improves macro recall, increasing from 42% to 56% (a relative improvement of 33.3% = (56% - 42%) / 42%), without any reduction in macro precision. This improvement is particularly pronounced for 2-, 3-, and 4-point reviews (referred to as non-extreme cases), where the relative recall improvement is 81.8% = (20% - 11%) / 11%, 200% = (45% - 15%) / 15%, and 138.8% = (43% - 18%) / 18%, respectively. These results suggest that embedding-based similarity is far more accurate at distinguishing non-extreme cases (i.e., all reviews except for 5-point perfect reviews and 1-point nonsensical reviews) than BLEU.

Fig. 7 shows the distribution of embedding-based similarity for each quality level. Compared to Fig. 4, we observe a substantial reduction in the overlap between the distributions. In Fig. 4, BLEU scores are heavily concentrated at the lower end, leading to significant overlap across different quality levels. In contrast, the embedding-based similarity distribution, as illustrated in Fig. 7, is more spread out, which helps reduce the overlap among quality levels.

## 6.5 RQ4: LLM-based Scoring

To evaluate the LLM-based review scoring approach proposed in Section 5.2, we applied it to the GradedReviews benchmark and compared the resulting scores (referred to as LLM scores) with those given by human experts (i.e., the reference scores in the benchmark). Our evaluation indicates a moderate positive correlation between LLM scores and human scores, with a Spearman Rank Correlation Coefficient of 0.47(ChatGPT)/0.49(DeepSeekCoder) and a p-value $\ll 0.05$. To facilitate comparisons across different performance metrics, we present the Spearman Rank Correlation Coefficients between each metric and the reference scores in Table 5. From this table, we observe that the p-values are consistently well below 0.05, indicating that the correlations shown in Table 5 are statistically significant. More importantly, the results show that LLM scores exhibit the strongest correlation with human scores, embedding-based similarity ranks in the middle, and MENTER has the weakest correlation. These findings suggest that LLM scores are the most sensitive to human evaluations and, consequently, more accurate than BLEU,MENTER, ROUGE and embedding-based similarity in assessing the quality of generated code reviews.

Table 5. Strength of Correlation between Performance Metrics and Human Scores

| Performance Metrics | Spearman Rank Correlation Coefficient | P-value |
|---|---|---|
| MENTER | 0.21 | 3.24e-52 |
| BLEU | 0.22 | 1.69e-59 |
| ROUGE | 0.25 | 1.89e-75 |
| Embedding-based Similarity | 0.38 | 3.19e-179 |
| LLM(ChatGPT-4o) Scores | 0.47 | 1.82e-280 |
| LLM(DeepSeek-Coder) Scores | 0.49 | 3.91e-305 |

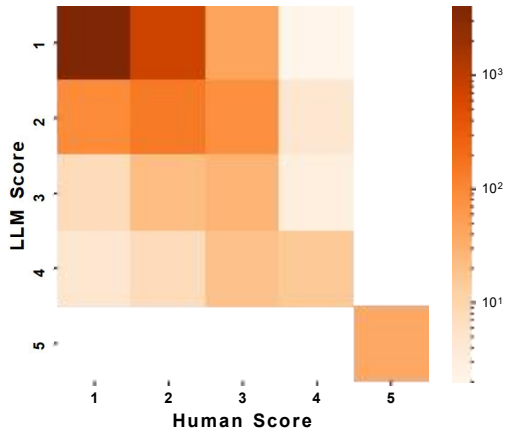

Fig. 8. Heatmap of Overlap Between LLM Scores and Human Scores

To further validate the consistency between LLM scores and human scores, we present a heatmap of their overlap in Fig.8 and Fig.9. From these figures, we observe that the data distribution is primarily concentrated along the diagonal. For instance, the hottest area is in the top-left corner, where both LLM scores and human scores are equal to one point. Notably, LLM scores are always equal to their corresponding human scores in all blocks along the diagonal. This concentration of hot areas along the diagonal suggests that LLM scores often match or closely align with human scores. Specifically, our calculations indicate that LLM scores have an 80.11% chance (4,137 out of 5,164 instances) of being exactly equivalent to human scores.

Good performance metrics should exhibit distinctly different distributions for low-quality and high-quality reviews, enabling effective differentiation between excellent and poor reviews. To assess this, we use the Kolmogorov-Smirnov (K-S) statistic [9] to evaluate whether the distributions of various metrics (e.g., BLEU scores, embedding-based similarity, and LLM scores) significantly differ across reviews of different quality levels. The evaluation results are presented in Table 6. Each cell in the table shows the K-S statistic for BLEU scores, embedding-based similarity, and LLM scores. Notably, the K-S statistic is computed pairwise, and the results are direction-independent, i.e., K-S(A,B) equals K-S(B,A). We also omit K-S(A,A) since it always equals zero, indicating no difference between the data sequence A and itself. From Table 6, we observe the following:

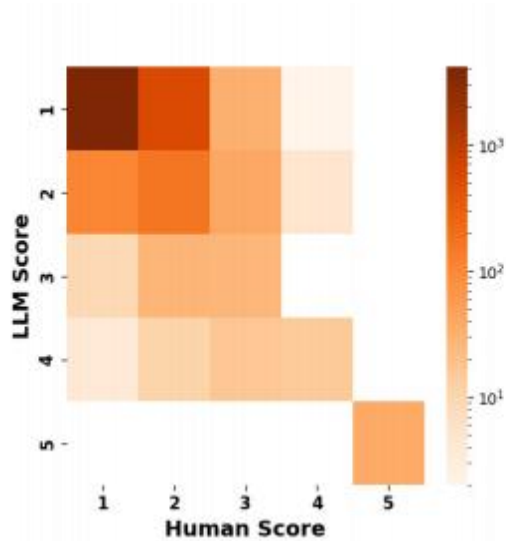

Fig. 9. Heatmap of Overlap Between LLM(deepSeek) Scores and Human Scores

- **Distinguishing Perfect Reviews:** All performance metrics are effective in distinguishing 5-point (perfect) reviews from other reviews. The K-S statistic between 5-point reviews and other reviews ranges from 0.97 to 1, indicating that the distribution of 5-point reviews is entirely (or nearly entirely) distinct from the distributions of other reviews.

- **Inefficiency BLEU:** Regarding K-S statistics, BLEU performs significantly worse than both embedding-based similarity and LLM scores. The K-S statistic for BLEU is consistently lower than that for embedding-based similarity and LLM scores, except for 5-point reviews. For instance, the K-S statistic between 1-point and 2-point reviews is 0.26, 0.55, and 0.55 for BLEU, embedding-based similarity, and LLM scores, respectively.

- **Comparison Between Embedding-Based Similarity and LLM Scores:** There is no significant difference between embedding-based similarity and LLM scores in terms of K-S statistics.

We conclude based on the preceding analysis that LLM scores are often more accurate than BLEU and embedding-based similarity in measuring the quality of code reviews against their references.

## 6.6 Threats to Validity

The first threat to external validity is that the benchmark employed the empirical study is composed of only 6,455 code reviews (including 5,164 generated reviews). The conclusions drawn on this benchmark are not necessarily true on other datasets. To mitigate the threat, we randomly select code reviews from one of the well-recognized high-quality dataset and generate code reviews with four state-of-the-art approaches. Another threat to external validity is that embedding models and LLMs are evolving quickly and thus the conclusions drawn on the selected embedding models and LLMs may not hold on new embedding models or LLMs in the near future.

A threat to construction validity is that the manual scoring requested by the benchmark construction could be subjective and inaccurate. To mitigate the threat, we request three experts to score each review and to reach a consensus on each review. Such experts have more than three years of programming and reviewing experience. As introduced in Section 4.3, we also employ an iterative process so that scorers can develop a common knowledge of scoring in the early stage of the manual scoring. The high Fleiss' Kappa coefficients as presented in Section 4.3 also suggest that there is a high consistency among the scorers.

Table 6. Kolmogorov-Smirnov Statistic Across Reviews of Varying Quality Levels

| Quality Level | 2-Point | 3-Point | 4-Point | 5-Point |
|---|---|---|---|---|
| 1-Point | BLEU: 0.26<br>Embedding: 0.55<br>LLM: 0.55<br>DeepSeek: 0.53 | BLEU: 0.36<br>Embedding: 0.71<br>LLM: 0.71<br>DeepSeek: 0.73 | BLEU: 0.44<br>Embedding: 0.73<br>LLM: 0.73<br>DeepSeek: 0.79 | BLEU: 1.00<br>Embedding: 1.00<br>LLM: 1.00<br>DeepSeek: 1.00 |
| 2-Point | | BLEU: 0.16<br>Embedding: 0.26<br>LLM: 0.24<br>DeepSeek: 0.28 | BLEU: 0.30<br>Embedding: 0.35<br>LLM: 0.47<br>DeepSeek: 0.55 | BLEU: 0.99<br>Embedding: 1.00<br>LLM: 1.00<br>DeepSeek: 1.00 |
| 3-Point | | | BLEU: 0.20<br>Embedding: 0.22<br>LLM: 0.28<br>DeepSeek: 0.33 | BLEU: 1.00<br>Embedding: 1.00<br>LLM: 1.00<br>DeepSeek: 1.00 |
| 4-Point | | | | BLEU: 0.97<br>Embedding: 1.00<br>LLM: 1.00<br>DeepSeek: 1.00 |

## 7 Discussions

The proposed approaches have the potential to be applied to various software engineering tasks. We notice that BLEU has been widely used for various tasks, e.g., code generation [14], source code migration (translation), code summary [32], commit message generation, natural language translation [30], and automatic text summarization [7]. We notice that in such tasks, BLEU has been employed to compute the lexical similarity between the generated text/code to its reference. However, for human beings, the semantic similarity between the generated texts/code and its reference is often more critical than lexicon similarity. What they really care is whether the generated code summary is semantically equivalent to the reference summary and whether the generated source code is semantically equivalent to the reference source code. However, two lexically different code summaries (or source code) could be semantically equivalent, but lexicon-based metrics like BLEU cannot identify their equivalence. In this case, the semantics-based approaches proposed in this paper could be employed to assess the generated text/code. We take it as a future work to validate the usefulness of the proposed approaches in such tasks.

Although both ChatGPT-4o and text-embedding-3-large were released recently by the same company (OpenAI), as explained in Section 6 the ChatGPT-based approach is much more accurate than the embedding-based approach (even with text-embedding-3-large as the underlying model). It may suggest that LLM-based direct comparison of text is preferable to converting it into an intermediate representation (embedding) before comparison. Consequently, it is potentially fruitful to replace the embedding-based comparison with LLM-based direct text/code comparison in various tasks, e.g., text retrieval, code search, clone detection, recommendation systems, and text clustering. Note that direct comparison could be much more expensive because it relies on pairwise comparison and each comparison requests a separate LLM query. In contrast, embedding-based approaches convert each text/code into a vector with a separate invocation of the embedding model, and compute the similarity automatically. As a result, the complexity of direct comparison is $o(n^2)$ whereas the complexity of embedding-based comparison is $o(n)$ where n is the number of text/code. That is one of the reasons why we present both LLM-based Scoring and embeddingSim in this paper: each of them has its own strengths and weaknesses.

It remains unclear how to take full advantage of various embedding models. We tried to combine various embedding models with various measures, e.g., taking the average/median/max/min of their embedding-based similarity. However, none of the combinations can further improve the performance of embedding-based scoring.

The size of the benchmark GradedReviews is limited. Noting that the benchmark construction requests tedious and resource-consuming manual scoring, we have to limit the size of the benchmark to control the cost. In total, it took around three man-month to manually score the reviews in the benchmark. In future, however, enlarging the benchmark could be significantly beneficial.

## 8   Conclusions and Future Work

Automated code review generation is highly desirable and new approaches in this line are emerging. However, the automated assessment of such approaches is less investigated. To this end, in this paper, we construct a benchmark, called GradedReviews for the automated assessment of code review generation approaches. We also propose new performance metrics, called embedding-based similarity, as well as new approach, called LLM scores, to measure the quality of automatically generated code reviews by comparing them against their reference reviews in the benchmark. A significant difference between the metrics/approaches proposed in this paper from existing metrics is that the former exploits the semantics of the reviews whereas the latter exploits their lexicon similarity. Our evaluation results suggest that such new metrics and new approaches are more accurate than the widely used BLEU. To the best of our knowledge, this paper is the first that focuses on the automated assessment of code review generation approaches.

In future, we would like to enlarge the benchmark by including code reviews generated by additional approaches/tools, especially new approaches proposed in the near future. We also plan to investigate whether and how the proposed metrics/approach could be employed to replace (or enhance) BLEU in measuring the quality of generated text/code besides code reviews, e.g., automatically generated code summary and commit messages.

## 9   Data Availability

The replication package and the benchmark are publicly available at [2].